 \documentclass[preprint2]{aastex}

\newcommand{\vdag}{(v)^\dagger}

\slugcomment{Not to appear in Nonlearned J., 45.}

\shorttitle{Desaturation of SDO/AIA images}
\shortauthors{Schwartz et al.}

\begin{document}


\title{A Systematic Approach to the Reconstruction of \\ Saturated {\em{SDO/AIA}} Images}


\author{R. A. Schwartz}
\affil{NASA Goddard Space Flight Center and Catholic University of America, Greenbelt, MD 20771 USA}
\email{richard.a.schwartz@nasa.gov}

\author{G. Torre}
\affil{Dipartimento di Matematica, Universit\`a di Genova and CNR - SPIN, Genova, via Dodecaneso 35 16146 Genova, Italy}
\email{torre@dima.unige.it}

\author{M. Piana}
\affil{Dipartimento di Matematica, Universit\`a di Genova and CNR - SPIN, Genova, via Dodecaneso 35 16146 Genova, Italy}
\email{piana@dima.unige.it}

%
%
%
%

%

\begin{abstract}
EUV images of solar flares provided by the {\em{Atmospheric Imaging Assembly}} onboard the {\em{Solar Dynamics Observatory}} ({\em{SDO/AIA}}) are often affected by saturation effects in their core, physically most interesting region. We introduce an image reconstruction procedure that allows recovering information in the primary saturation domain using the secondary images produced by the diffraction fringes as input data. Such a procedure is based on standard image-processing tools like correlation, convolution, and back-projection. Its effectiveness is tested in the case of {\em{SDO/AIA}} observations of the July 8 2013 flaring event.
\end{abstract}





\section{Introduction}
Since its launch occurred on February 11 2010, the {\em{Atmospheric Imaging Assembly}} in the {\em{Solar Dynamics Observatory (SDO/AIA)}} has been contributing to our knowledge of solar flare emission, permitting full disk imaging in $10$ wavebands (seven of which are centered at EUV wavelengths), at a spatial resolution of about $1$ arcsec and with a temporal cadence of $12$ sec in each of the EUV bands. However, the scientific potential of {\em{SDO/AIA}} images is limited by the presence of significant saturation, which flattens and muddles the brightest, physically most interesting core of such images for even relatively modest flares. 

{\em{Saturation}} refers to the condition where,  in a CCD-based imaging system, pixels lose their ability to accommodate additional charge ({\em{primary saturation}}), causing charge to spill into their neighbors. This secondary effect, named {\em{blooming}},  also induces bright rays typically along the north-south (for AIA) axis in the image. 

We now introduce a computational method that utilizes the diffraction component of the Point Spread Function (PSF) of {\em{SDO/AIA}} \citep{boetal12,poetal13,grsuwe11} to recover the signal in the primary saturation region. The PSF of each {\em{AIA}} telescope models the system response to a point source and describes blurring, dispersion, and diffraction. The various {\em{AIA}} PSFs depend on the wavelength of the incoming photons and present a two-component structure which is similar to those of the {\em{TRACE}} passbands \citep{gbsyma06}. The main idea of our approach is to exploit,  in particular, the diffraction pattern showing up {\em{outside}} the core of the saturated image to recover information {\em{inside}} the saturated region. The feasibility of this approach was demonstrated by previous analyses of {\em{TRACE}} and {\em{AIA}} diffraction patterns \citep{rakrli11,linita01,grsuwe11} where, however, the inverse diffraction problem is addressed by means of non-automatic, laborious and semi-heuristic procedures. 

Our approach is based on three image-processing steps. In the first step ({\em{correlation}}) we identify the region of primary saturation distinguishing it from blooming on the basis of a data-dependent correlation procedure. In the second step ({\em{convolution}}) the diffraction fringes in the image are localized by convolving the diffraction PSF with the pixels in the region of primary saturation. The third step ({\em{back-projection}}) is used to reconstruct the true image in the primary saturation region using the information contained in the diffracted images. Details about these steps are given in Section 2. Section 3 illustrates the performances of the method in the case of {\em{SDO/AIA}} data recorded during the July 8 2013 event. Our conclusions are offered in Section 4.
 
\section{The reconstruction method}
\begin{figure}[pht]
\includegraphics[width=0.5\textwidth]{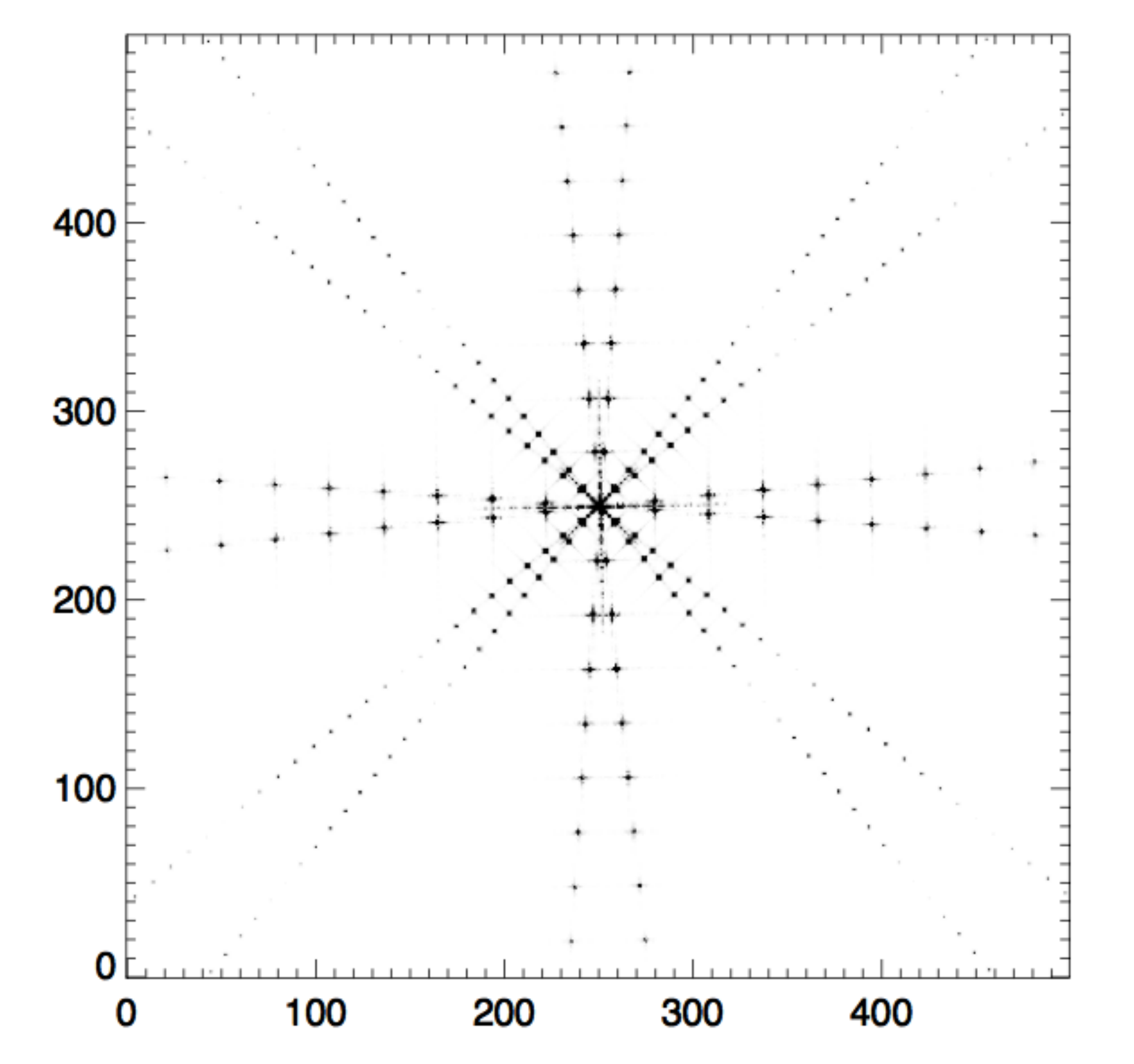} 
\caption{The {\em{SDO/AIA}} Point Spread Function. The $304~ \AA$ PSF is superimposed to the $131~ \AA$ PSF and rotated of $44$ deg. A threshold has been applied in order to show the diffraction patterns over the full range of the FOV.}
\label{fig:psf}
\end{figure}

The general model equation mimicking the signal formation process in an optical system is
\begin{equation}\label{general}
g = K * f + b,
\end{equation}
where $g$ is the measured data, $K$ is the global PSF, $f$ is the true flux distribution, and $b$ is the background. This equation applies to all points in the data image. However, in the case of the {\em{SDO/AIA}} image reconstruction problem the use of this equation should account for the fact that
\begin{itemize}
\item the true image can be recovered just in a sub-domain of the Field Of View (FOV) corresponding to the primary saturation region;
\item since the total diffracted image is very intense, the usable data, $g$, is just the one containing the diffraction pattern;
\item just the diffraction component of the PSF is useful for this operation.
\end{itemize}
In fact, the PSF associated with each {\em{SDO/AIA}} wavelength can be realistically modeled as
\begin{equation}\label{psf-components}
K(x,y) = K_C(x,y) + K_D(x,y)
\end{equation}
where $K_C(x,y)$ is the core, Gaussian-like component mimicking diffusion, and $K_D(x,y)$ is the diffraction component wherein about $20\%$ of the total flux is coherently scattered due to the two square wire grid meshes supporting the thin EUV filters. Figure \ref{fig:psf} shows the PSF of the $131~ \AA$ passband with, superimposed and rotated of $44$ deg, the one of the $304 ~\AA$ passband (we applied a threshold in order to show the diffraction patterns over the full range of the FOV). The figure points out the presence of 8 arms of diffraction for each passband, how diffraction scales with wavelength and how dispersion in the diffraction peaks scales with respect to dispersion in the core.

\begin{figure*}[pht]
\begin{center}
\begin{tabular}{cc}
\includegraphics[width=6.cm]{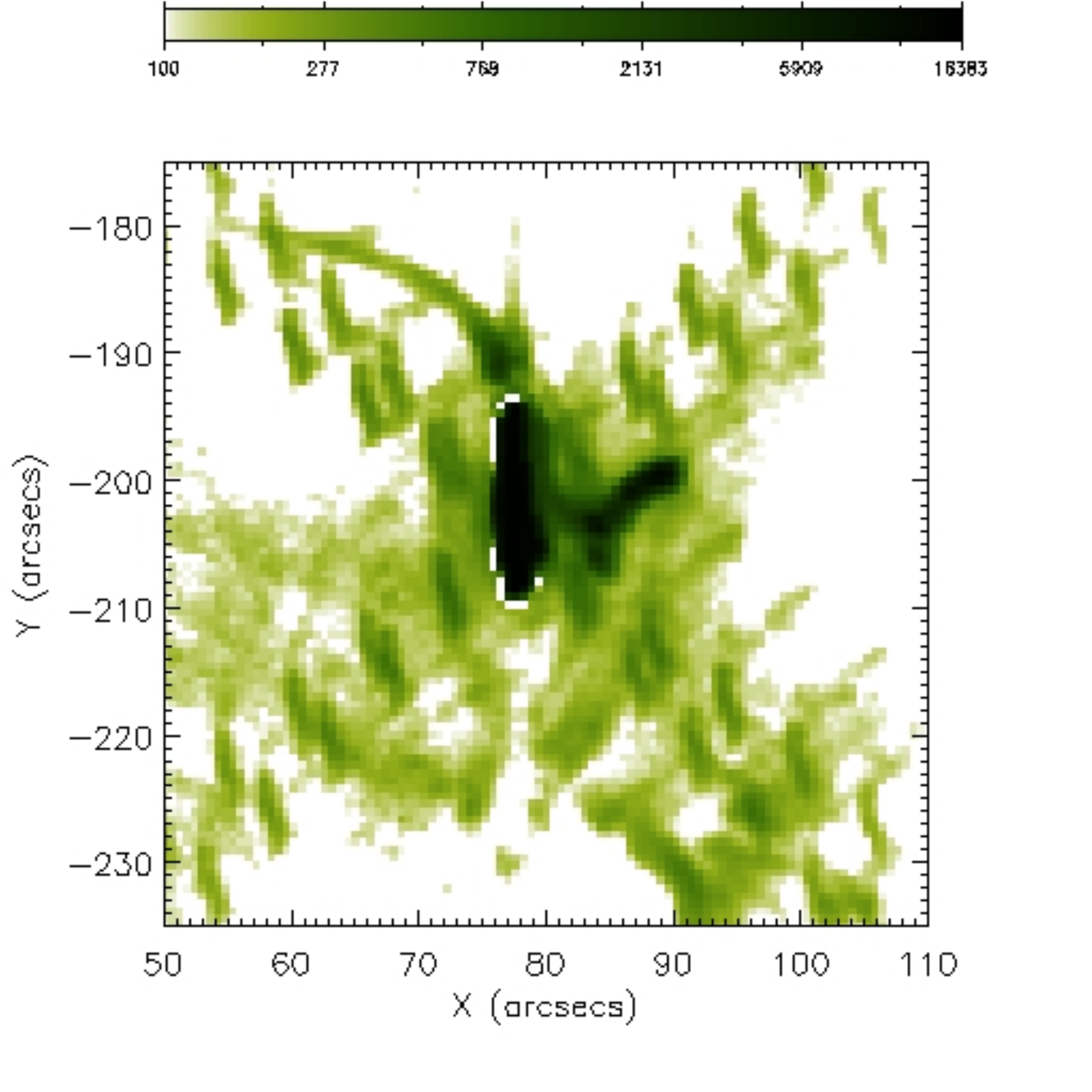} &
\includegraphics[width=6.cm]{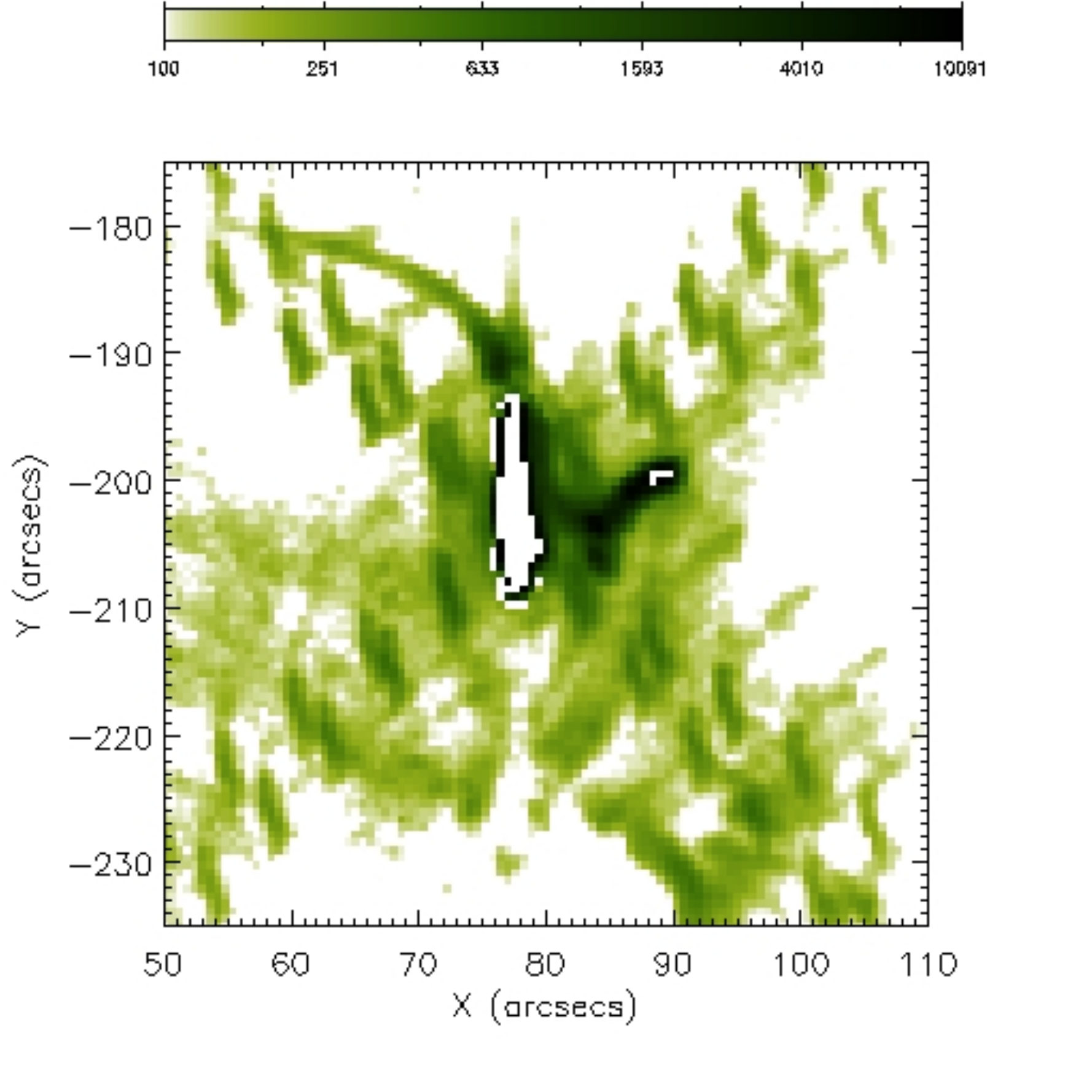} \\
\includegraphics[width=6.cm]{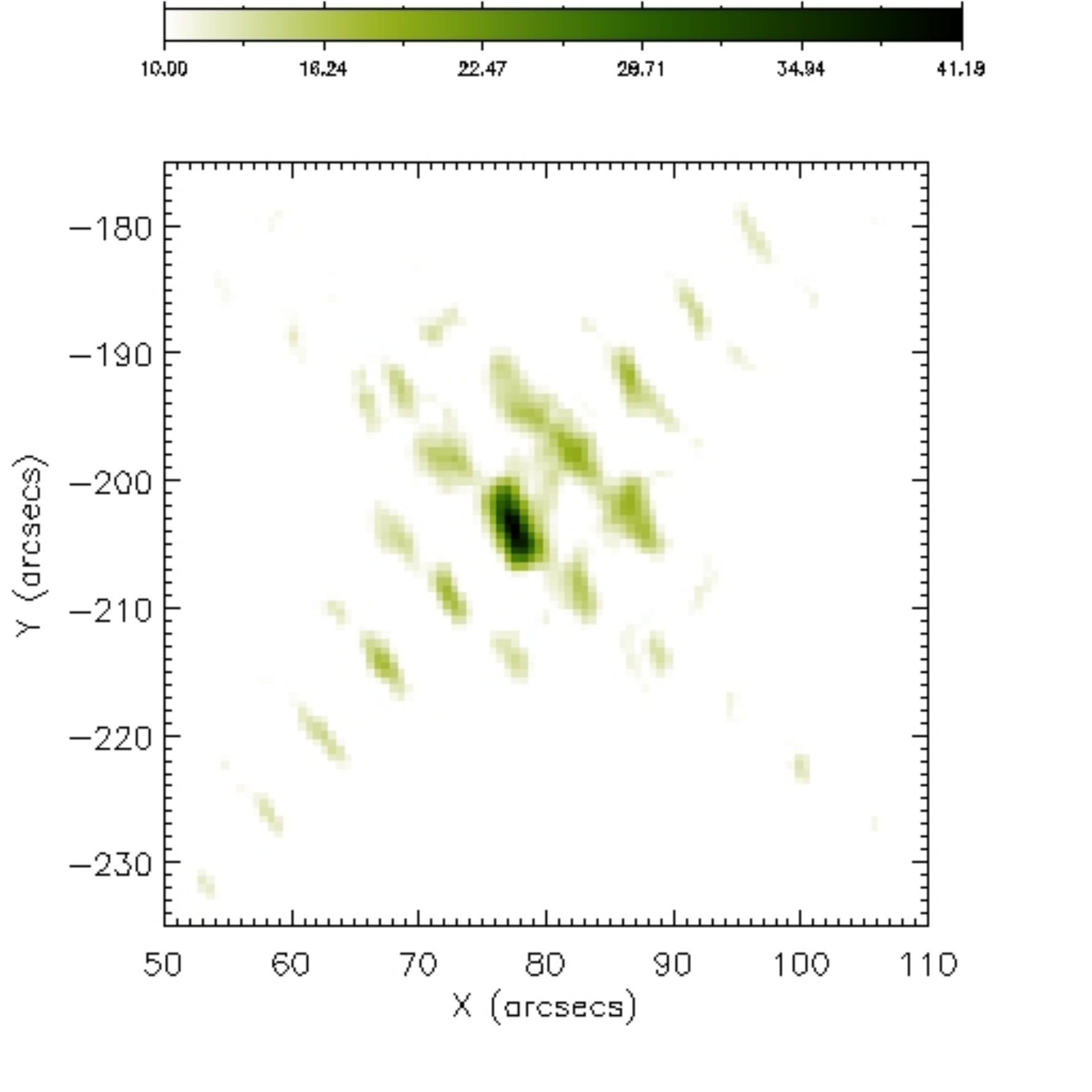} &
\includegraphics[width=6.cm]{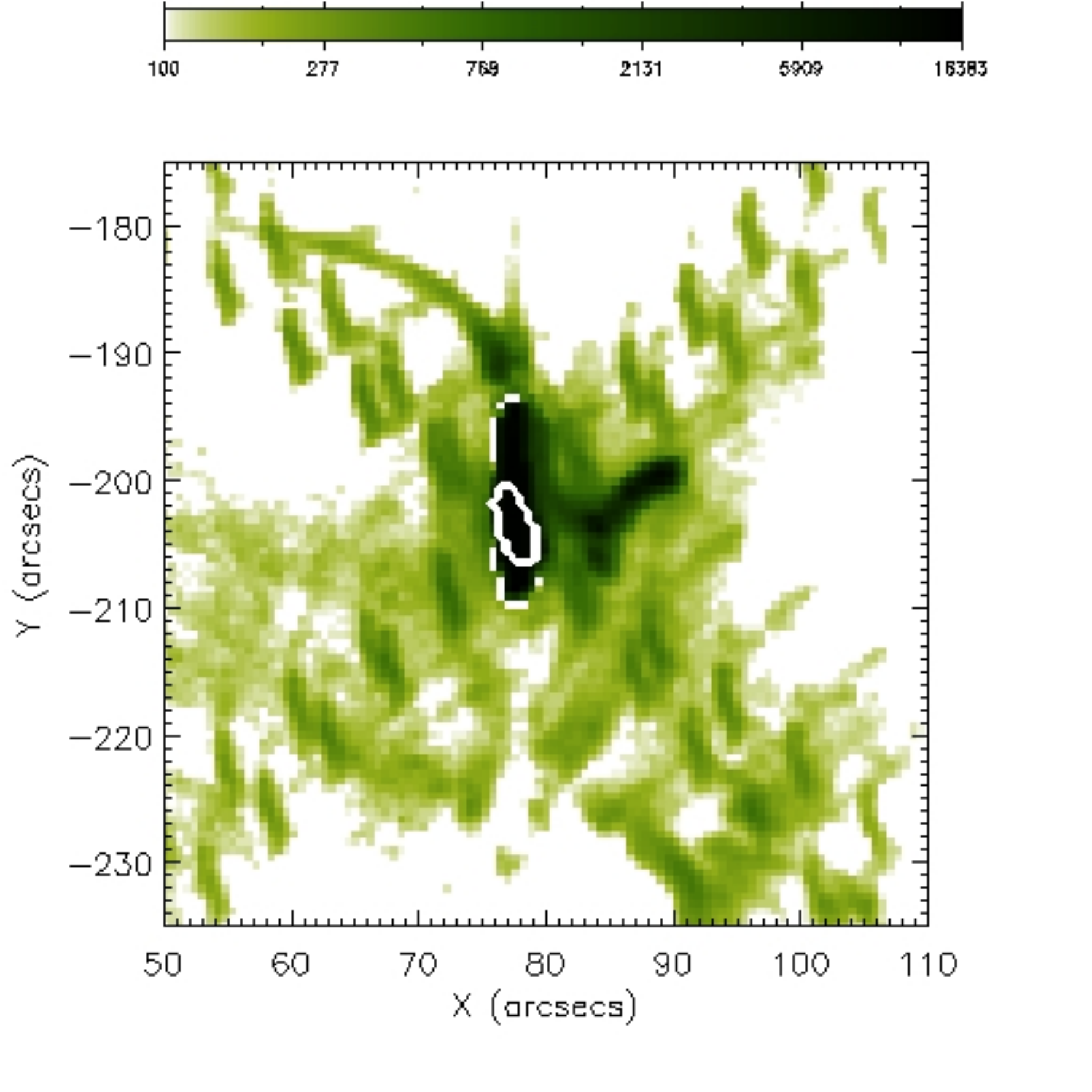} \\
\includegraphics[width=6.cm]{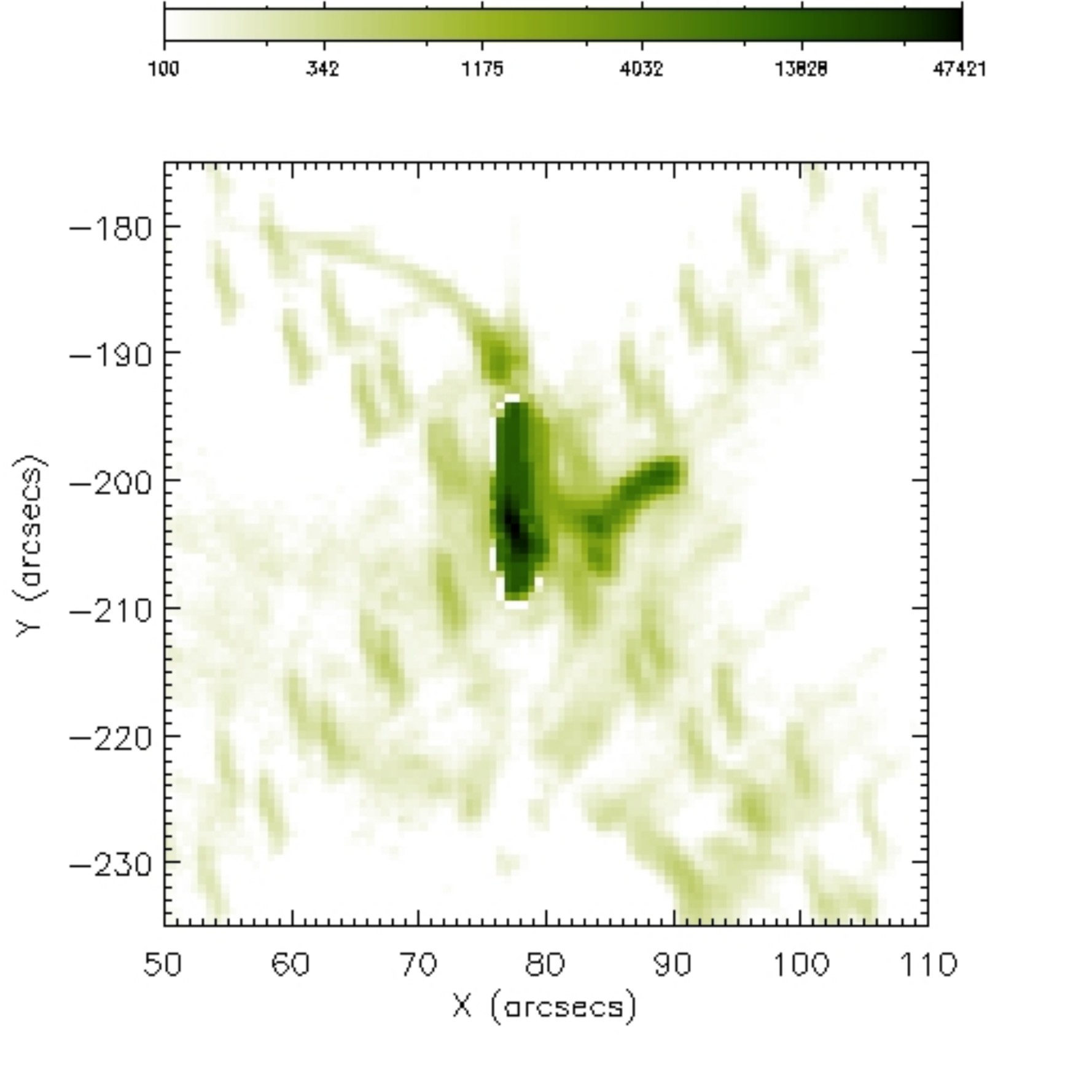} &
\includegraphics[width=6.cm]{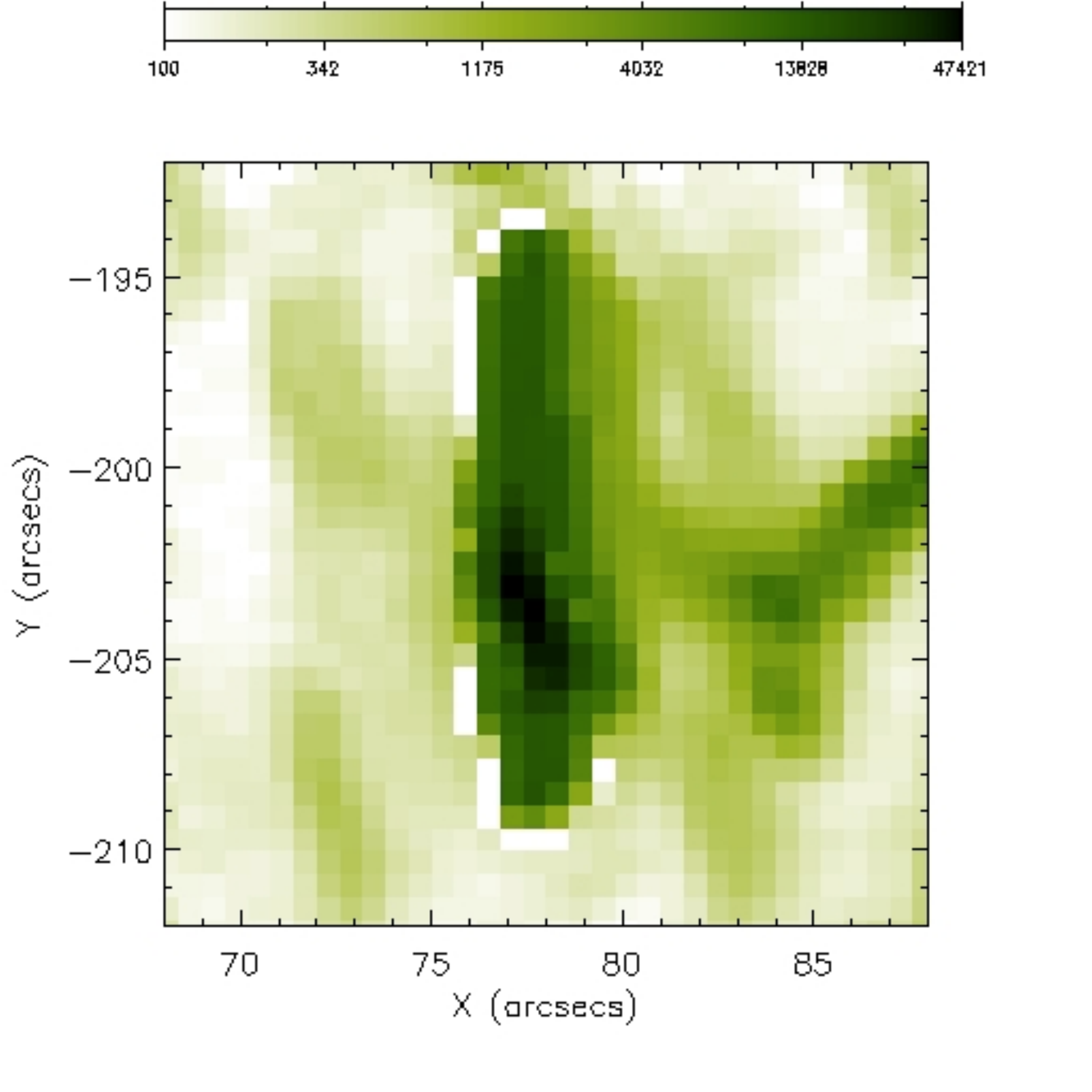} 
\end{tabular}
\caption{From a saturated to a de-saturated {\em{AIA}} image using correlation/reconstruction. All maps are log-scaled and the green-white linear color table is used for their representation. Top left panel: July 8 2013 event at 01:21:32 UT, observed at $131~ \AA$ (the white artifact around the left side of the saturation region was present in the original data and may be due to the pre-processing step; it has been left untouched by our analysis). Top right panel: the same image but with the saturation region masked. Middle left panel: representation of the correlation between the diffraction PSF and the masked image. Middle right panel: the primary saturation region is bounded  by the white contour in the observed image. Bottom left panel: the reconstruction offered by EM. Bottom right panel: zoom on the reconstructed core region.}
\label{fig:ingredients}
\end{center}
\end{figure*}

The diffraction component of the PSF plays a crucial role in the de-saturation process described in this Section. The workflow of such a process is illustrated in Figure \ref{fig:ingredients} for a flare occurring on July 8 2013, measured at $131~ \AA$ (more details on this event will be given in Section 3). The computational steps at the basis of its realization are:
\begin{description}
\item{\bf{Correlation.}} Let us denote with $S$ the overall saturation domain in the {\em{AIA}} image in the top left panel of Figure \ref{fig:ingredients}. $S$ is the sum of the primary saturation region $PS$ and the blooming region $B$ and all its pixels have intensity of around $16000$ DN pixel$^{-1}$. Such pixels are masked away to obtain the image $g^{\prime}$ represented in the top right panel of Figure \ref{fig:ingredients}. The back-projection matrix $K_D^T$ is the transpose of the diffraction matrix $K_D$ and has size equal to the number of unsaturated pixels times the number of saturated pixels. We computed the correlation map
\begin{equation}\label{correlation-map}
C = K_D^T g^{\prime}
\end{equation}
for all pixels in $S$, and show it in the middle left panel of Figure \ref{fig:ingredients}. Since primary saturation is identified by those pixels in $S$ for which $C$ is high, a thresholding procedure allowed the identification of $PS$ in the correlation map. 
\item{\bf{Convolution.}} The white curve in Figure \ref{fig:ingredients}, middle right panel, bounds the domain $PS$ in the observed saturated image where information can be recovered by means of image reconstruction. We computed the convolution between $PS$ and $K_D$ to localize the diffraction fringe in $g^{\prime}$ and utilized just that information, indicated with $g$ in the reconstruction procedure. In fact $g$ contains pixels where the diffraction effect is most significant.
\item{\bf{Back-projection.}} The reconstruction of information in the region of primary saturation $PS$ required the computation of the back-projection
\begin{equation}\label{back-projection}
f = K_D^T g
\end{equation}
where $f$ is defined in $PS$ and here $K_D^T$ has size equal to the number of pixels in the diffraction fringes times the number of pixels in the primary saturation region. The result is the de-saturated image at the bottom left panel of Figure \ref{fig:ingredients} where the blooming effects are still present, the diffraction pattern is clearly visible and the recovery of information in the region originally affected by primary saturation is evident. Such a recovery of information is even more evident in the last, bottom right panel of the figure, a zoom on the de-saturated domain. We note that the use of the log scale better shows the presence of the diffraction fringes with respect to the reconstructed signal.
\end{description}

From a computational viewpoint, the correlation map $C$ in (\ref{correlation-map}) and the back-projected map $f$ in (\ref{back-projection}) have been obtained by applying Expectation Maximization (EM) \citep{deetal77} to $g^{\prime}$ and $g$ respectively. 
EM (also known as the Lucy-Richardson method in this context \citep{lu74}) realizes maximum likelihood in the case of Poisson data and imposing positivity on the solution space (CCD data are more correctly quasi-Poisson but EM effectiveness relies more on positivity than on the way data-prediction discrepancy is computed). EM iteration in the back-projection case is 
\begin{equation}\label{EM}
f^{k+1} = \frac{f^{k}}{K_D^T I} K^T_D \left( \frac{g}{K_Df^k + b}\right),
\end{equation}
where $I$ is the unit vector and $b$ is an estimate of the background. In our implementation, the stopping rule that avoids numerical instability is based on the concept of constrained residual \citep{bepi14} and imposes that the empirical value of this random variable coincides with its expectation value \citep{beetal13}. We point out that for bright saturated images which are intrinsically small, diffraction places almost perfect copies over the fringes of the eight separate arms. In this case, the back-projection side-lobes are unambiguously removed and the formulation becomes a version of 2-D epoch-folding completely analogous to 1-D pulse synthesis \citep{leelwe83}.


\begin{figure*}[pht]
\begin{center}
\begin{tabular}{cc}
\includegraphics[width=6.3cm]{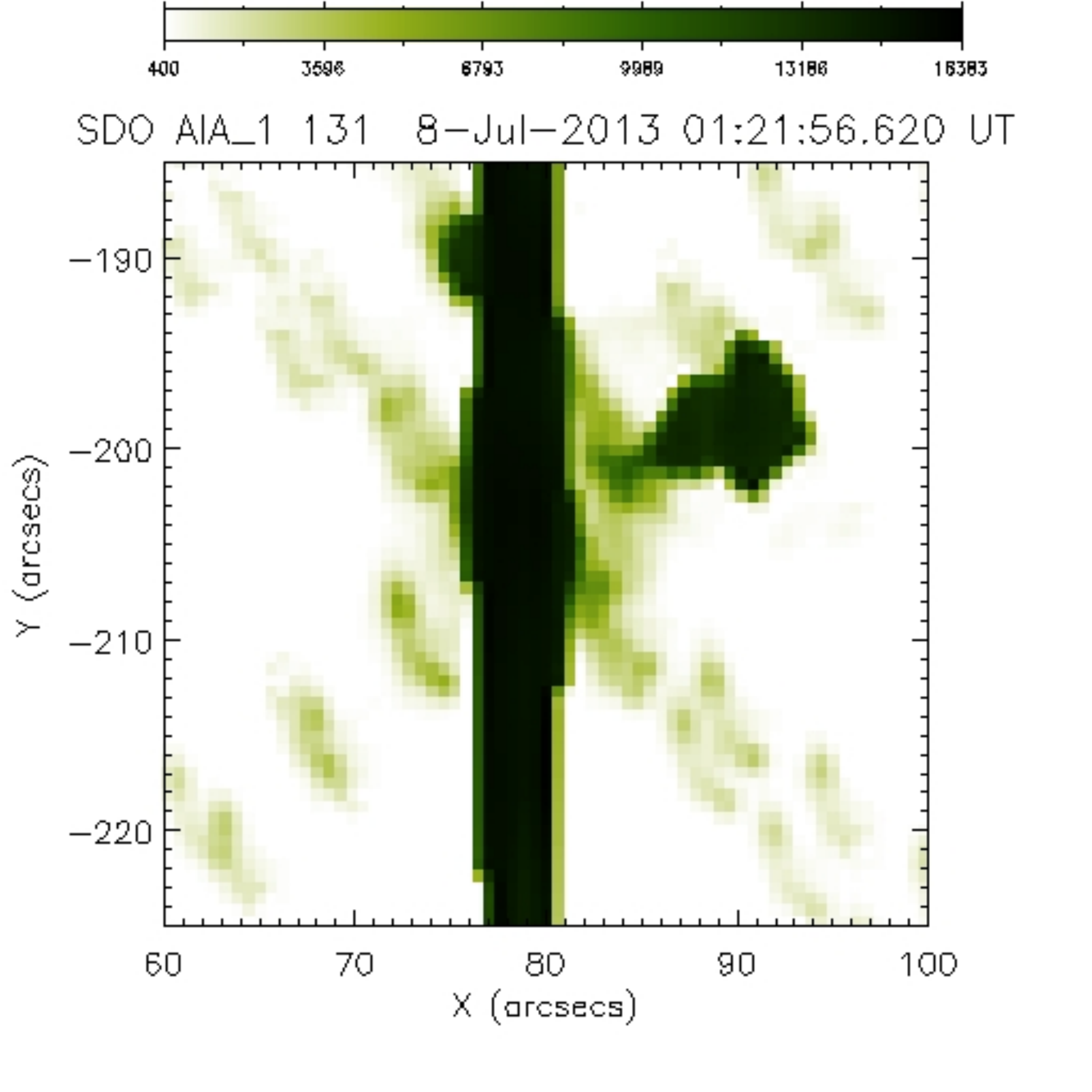} &
\includegraphics[width=6.3cm]{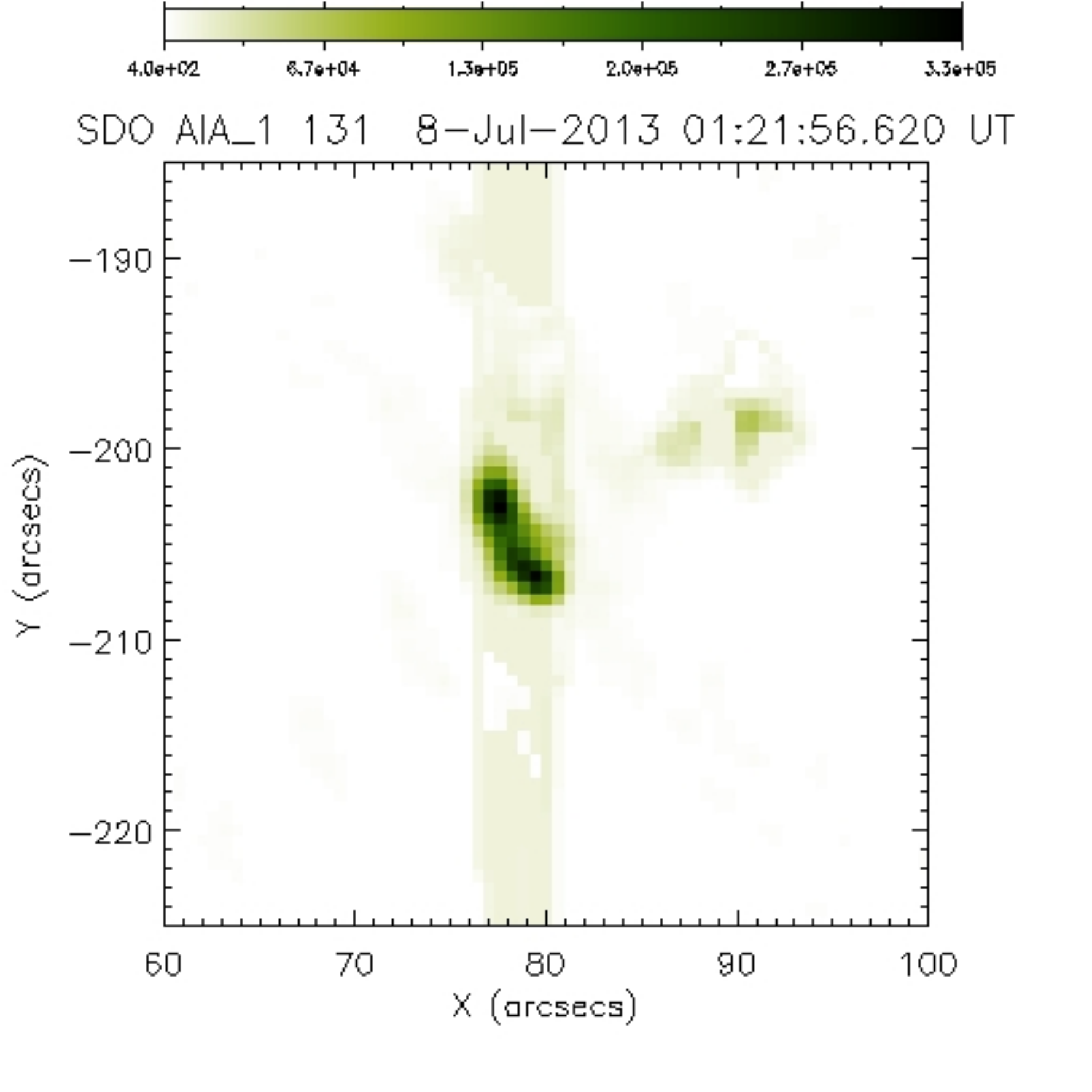} \\
\includegraphics[width=6.3cm]{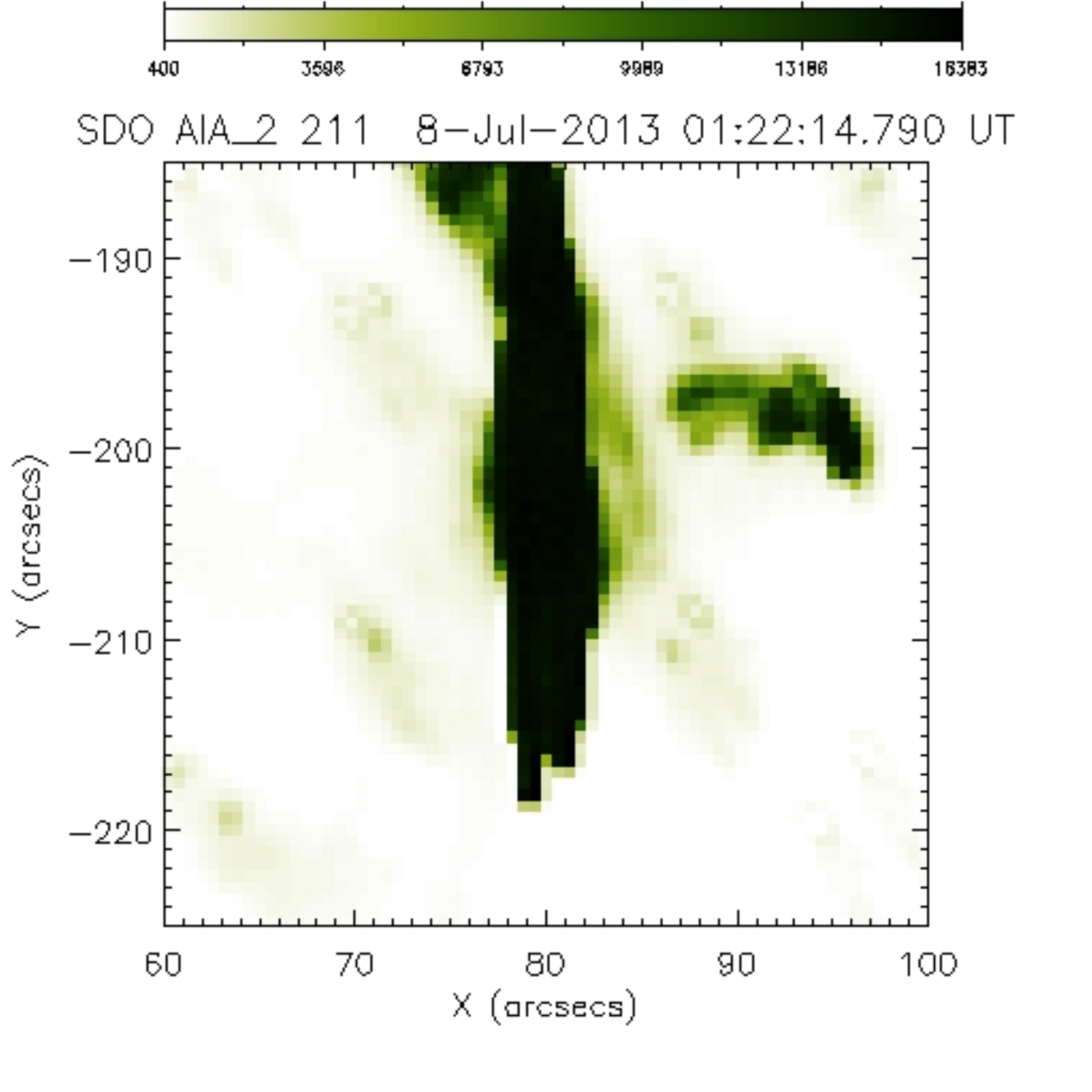} &
\includegraphics[width=6.3cm]{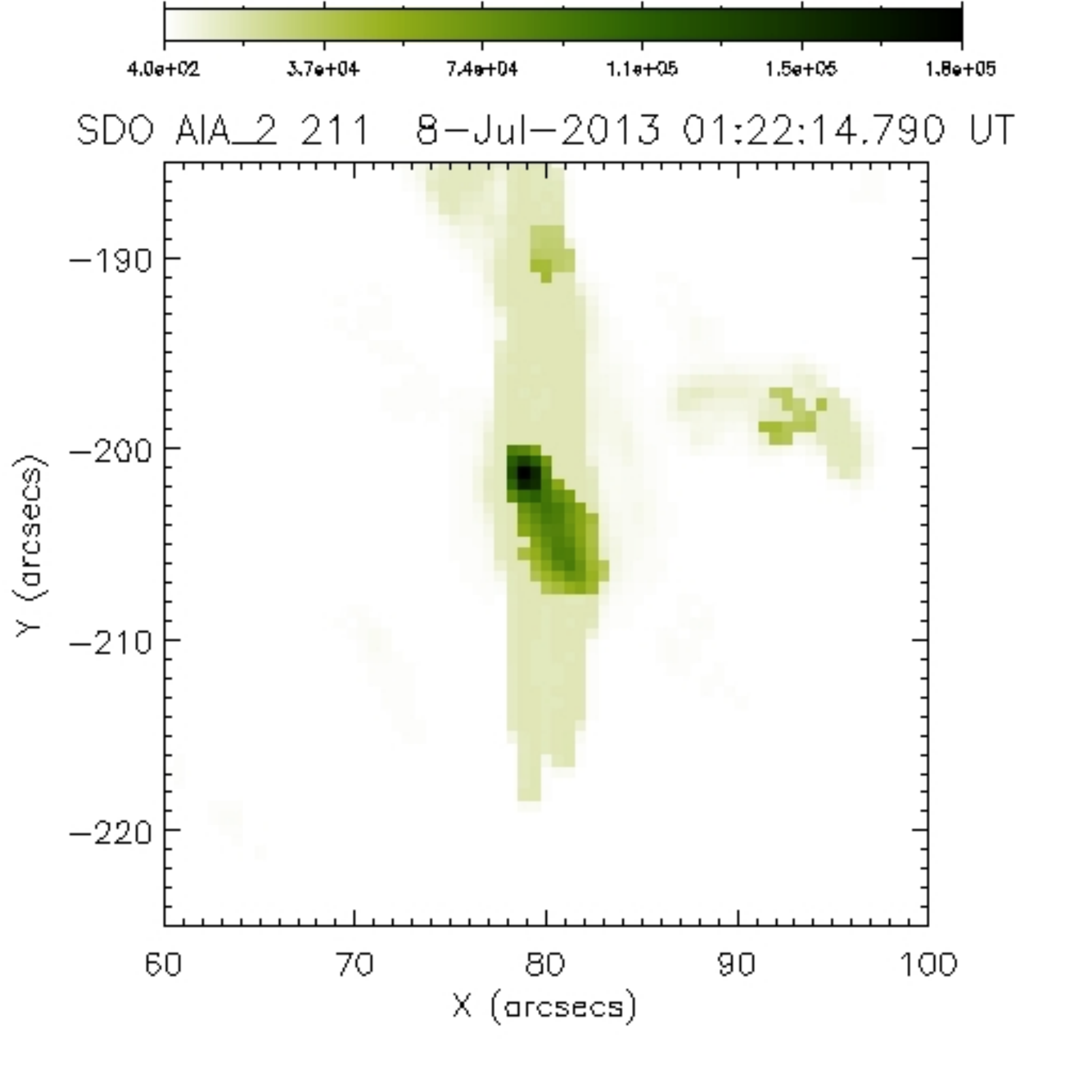} \\
\includegraphics[width=6.3cm]{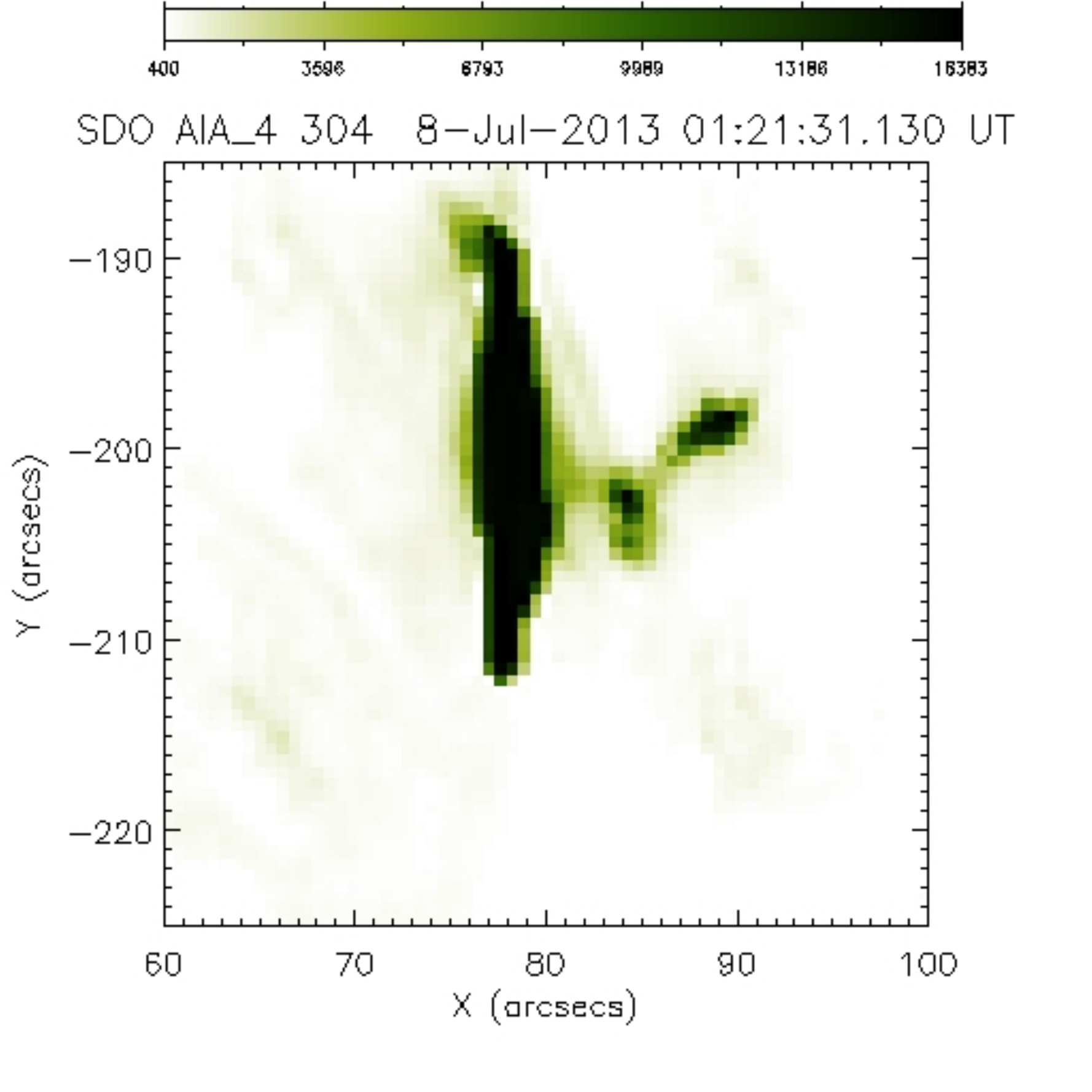} &
\includegraphics[width=6.3cm]{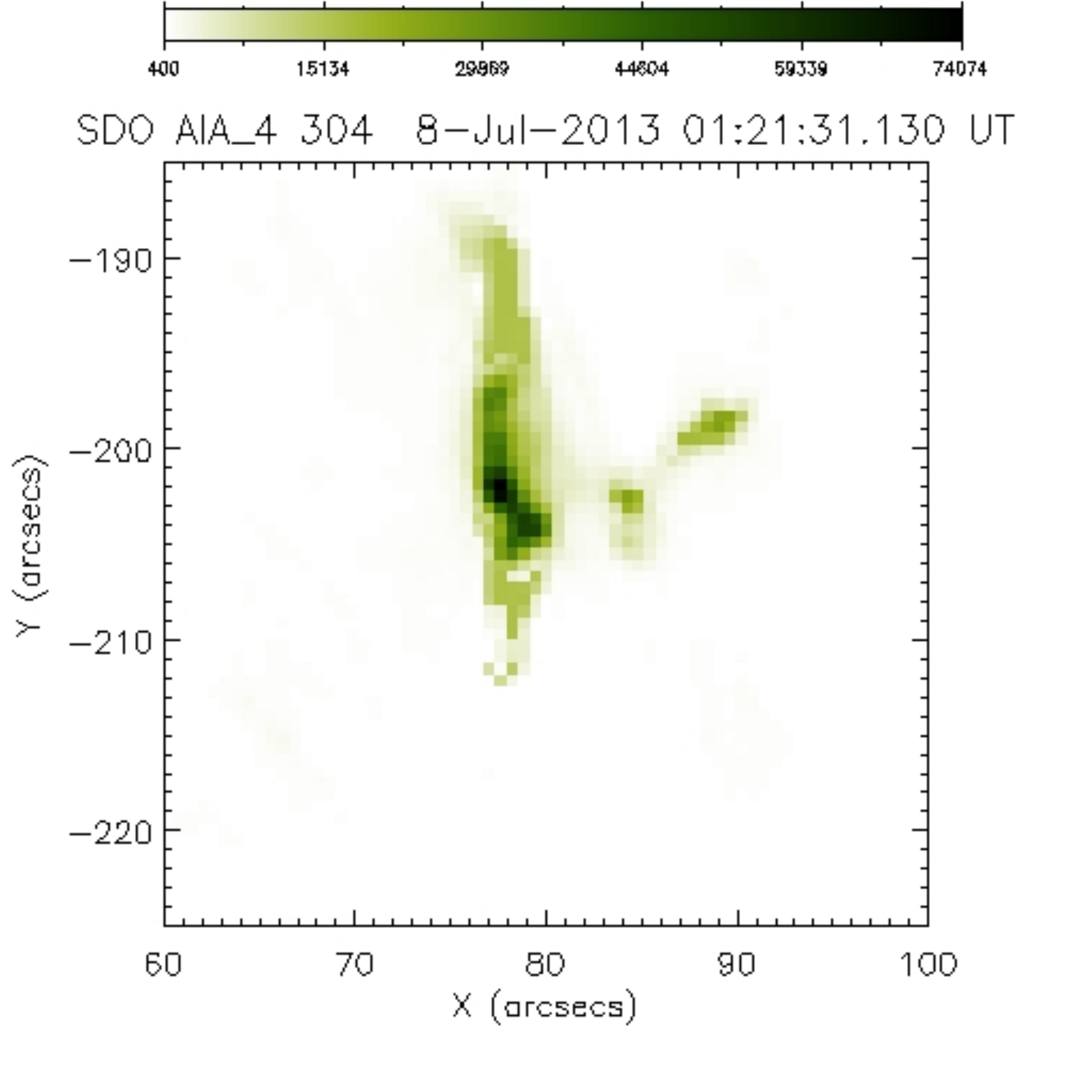} 
\end{tabular}
\caption{The de-saturation method at work. The event is the July 8 2013 one. The first row corresponds to data in the $131 ~\AA$ wavelength at times 01:21:56 UT; the second row corresponds to data in the $211~ \AA$ wavelength at times 01:22:14 UT; the third row corresponds to data in the $304~ \AA$ wavelength at times 01:21:31 UT. For each dataset, we provided the saturated image and the corresponding de-saturated image, both in linear scale.}
\label{fig:de-saturated}
\end{center}
\end{figure*}

\section{Applications}
On July 8 2013  the Sun experienced a C9.7 GOES class flare in the time range 01:13 - 01:23 UT with peak at around 01:22 UT. This event was very well observed by {\em{RHESSI}} but, despite the moderate peak flux, the corresponding {\em{AIA}} images saturated at most wavelengths and times. In principle, simultaneous solar flare observations with {\em{SDO}} and {\em{RHESSI}} provide spatially resolved information on hot plasma and energetic particles during flares and the unavailability of EUV maps due to saturation clearly limits this kind of analysis. As a test of our de-saturation approach we considered this event and, specifically, we applied the method to a number of saturated images collected at three different wavelengths ($131 ~\AA$, $211~ \AA$ and $304~ \AA$) and at two times for each wavelength. We downloaded {\em{AIA}} level 1.5 data from http://www.lmsal.com/get$\_$aia$\_$data and used a slightly modified version of the routine "aia$\_$calc$\_$psf.pro" in Solar SoftWare (SSW) for the generation of the diffraction PSF \citep{grsuwe11} (since our method strongly relies on the characteristics of the applied PSF, it would be interesting to run the reconstruction algorithm also in the case of the semi-empirical PSF described by \citet{poetal13}). The results of our analysis are given in Table \ref{table:c-stat} that reports the Cash statistic values predicted by the de-saturated images. Specifically, we have computed the C-statistic of a set made of $100$ random samples of the fringes, repeated this process $10$ times (every time changing the set of random samples) and reported the average values over these ten times (we point out that the C-statistic values may suffer because our computation of the PSF does not allow for an emission line distribution within the AIA passbands). Further, Figure \ref{fig:de-saturated} visually compares saturated and de-saturated maps corresponding to one of the two observation times for each wavelength. We point out that the linear scale softens the effect of diffraction fringes and blooming  (blooming cannot be eliminated by means of reconstruction but its visual effect could be further reduced by means of interpolation) . 

\begin{figure}[pht]
\begin{center}
\begin{tabular}{c}
\includegraphics[width=7.cm]{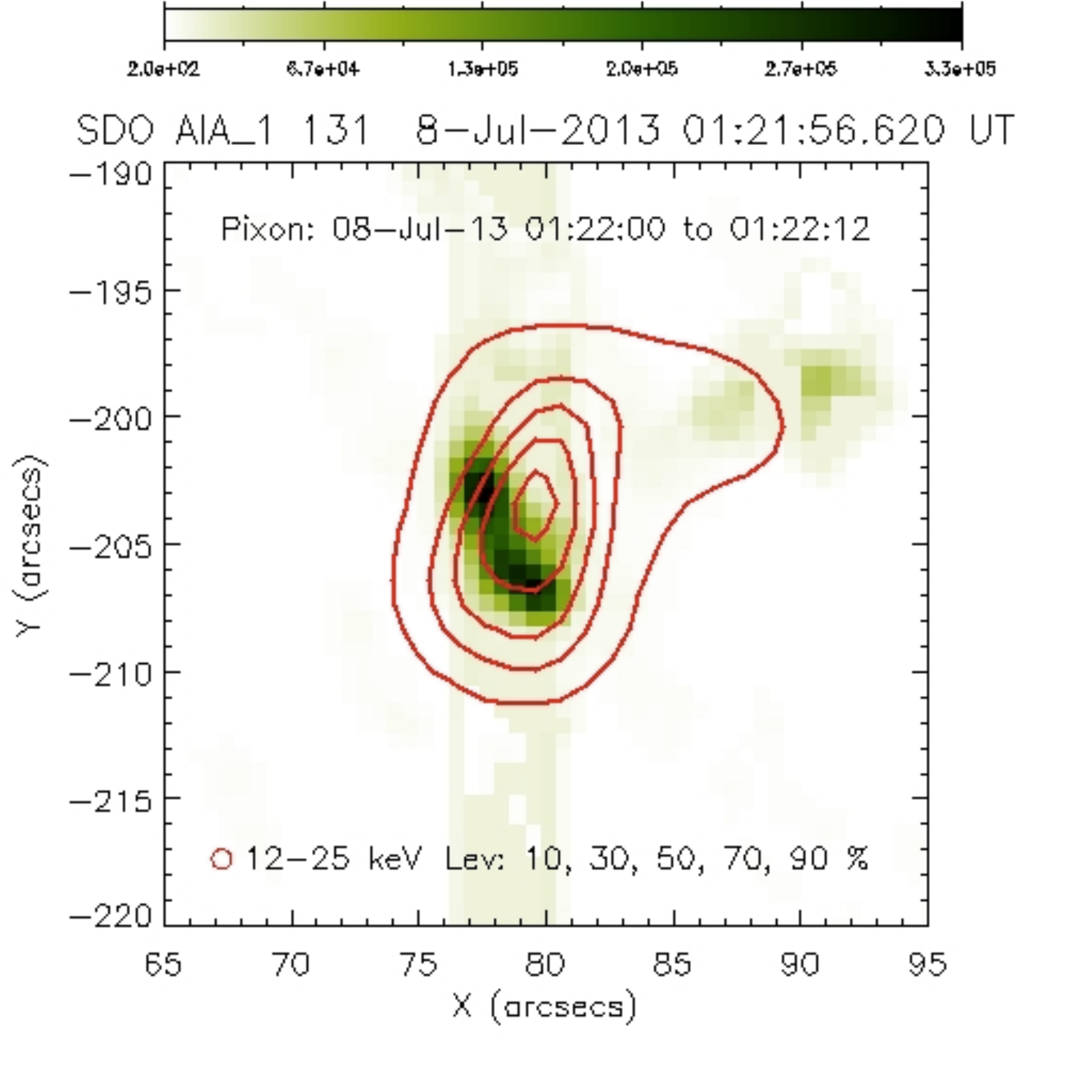}
\end{tabular}
\caption{Integration of {\em{RHESSI}} data and SDO/AIA de-saturated images in the case of the July 8 2013 event. Contours of a RHESSI image reconstructed by Pixon in the time interval 01:22:14 - 01:22:26 UT in the energy channel $12-25$ keV are superimposed on the de-saturated linear-scale AIA map acquired by the $131~ \AA$ telescope at 01:22:20 UT.}
\label{fig:RHESSI-AIA}
\end{center}
\end{figure}

The potential of this method for a more extensive use of EUV data is illustrated in the example of Figure \ref{fig:RHESSI-AIA}. To construct it, we first considered a {\em{RHESSI}} image of hard X-ray flux reconstructed by Pixon \citep{pugo05} in the time interval 01:22:14 - 01:22:26 UT for the energy channel $12-25$ keV. In the figure, the contours of this X-ray image are superimposed on the de-saturated $131 \AA$ image corresponding to the {\em{SDO/AIA}} observation occurred at 01:22:20 UT (an offset of around $2$ arcsec is visible  between the {\em{RHESSI}} and {\em{AIA}} images).

\begin{table}
\begin{center}
\begin{tabular}{c|c|c}
wavelength & time (UT) & C-stat \\
\hline \hline
 & 01:21:56 &   2.085 \\
 $131 \AA$ & 01.22:20  & 3.739 \\
\hline
 & 01:22:14 &  1.243 \\
 $ 211 \AA$ & 01:22:23 &  2.046 \\
 \hline
  & 01:21:31 & 2.275 \\
 $304 \AA$ & 01:21:43 & 2.796 \\
\end{tabular}
\end{center}
\caption{Cash statistic for SDO/AIA datasets acquired by three different wavebands. For each line, two acquisition times are considered.}
\label{table:c-stat}
\end{table}

\section{Conclusions}
The EUV analysis of flaring events by means of {\em{SDO/AIA}} is notably limited by saturation and blooming, occurring at all wavelengths and times even in the case of moderate peak flux. This Letter illustrates a method that addresses, for the first time in a systematic way, the problem of eliminating primary saturation from these maps, utilizing standard image processing techniques like correlation, convolution and statistical deconvolution (the problems of recovering information in the blooming region and removing the multiple diffraction components will be addressed in future research). We applied our approach to data recorded during the July 8 2013 flare at three different EUV wavelengths and showed its effectiveness to reduce saturation in a variety of artifact configurations and intensities. Once fully automated, these de-saturation procedures will be included in Solar SoftWare. Further we will be extending this technique to include the wavelength dependent dispersion of the PSF to integrate the method of \citet{rakrli11} into this schema.

The impact of this method on the exploitation of EUV information in solar flares can be notable, for example in imaging spectroscopy studies of the dynamic response of solar corona and chromosphere to solar flares. Another broad and significant applicability domain involves multi-spectral studies. For example, combined analysis of {\em{SDO/AIA}} and {\em{RHESSI}} data may provide information on crucial plasma parameters like differential emission measure, temperature and density, which would notably contribute to unveil still mysterious processes concerned with energy transport mechanisms in flares.

\begin{acknowledgments}
Richard Schwartz was supported by NASA grant NNX14AG06G. Gabriele Torre was supported during his stay at Goddard by The Catholic University of America in Washington, DC, using {\em{RHESSI}} funds under grant number NNX11AB37G. Anna Maria Massone and Federico Benvenuto are kindly acknowledged for fruitful discussions.
\end{acknowledgments}


\begin{thebibliography}{}
\bibitem[Benvenuto et al.(2013)]{beetal13} Benvenuto, F., Schwartz, R. A., Piana, M. \& Massone, A. M. 2013, \aap, 555, A61
\bibitem[Benvenuto and Piana(2014)]{bepi14} Benvenuto, F. \& Piana, M. 2014, Inverse Problems, 30 035012
\bibitem[Boerner et al.(2012)]{boetal12} Boerner, P., Edwards, C., Lemen, J., Rausch, A. Schrijver, C., Shine, R. et al 2012, \solphys, 275, 41
\bibitem[Dempster et al.(1977)]{deetal77} Dempster, A. P., Laird, N. M. \& Rubin, D. B. 1977, J. R. Stat. Soc. B, 39, 1
\bibitem[Gburek et al.(2006)]{gbsyma06} Gburek, S., Sylwester, J. \& Martens, P 2006, \solphys, 329,531
\bibitem[Grigis et al.(2011)]{grsuwe11} Grigis, P., Su, Y. \& Weber, M. 2011, AIA PSF Characterization and Image Deconvolution (http://www.lmsal,com/sdodocs/)
\bibitem[Hoegbom(1974)]{ho74} Hoegbom, J. A. 1974, \aap, 15, 417
\bibitem[Leahy et al(1983)]{leelwe83} Leahy, D. A., Elsner, R. F \& Weisskopf, M. M 1983, \apj, 272, 256
\bibitem[Lin et al.(2001)]{linita01} Lin, A. C., Nightingale, R. W. \& Tarbell, T. D. 2001, \solphys, 198, 385
\bibitem[Lucy(1974)]{lu74} Lucy, L. B. 1974, Astron. J., 79, 745
\bibitem[Poduval et al.(2013)]{poetal13} Poduval, B., DeForest, C. E., Schmelz, J. T. \& Pathak, S. 2013, \apj, 765, 144
\bibitem[Puetter and Gosnell(2006)]{pugo05} Puetter, R. C. \& Gosnell, T. R. 2006, Annu. Rev. Astron. Astrophys., 43, 139
\bibitem[Raftery et al.(2011)]{rakrli11} Raftery, C. L., Krucker, S. \& Lin, R. P. 2011, \apjl, 743, L27 


\end{thebibliography}
\end{document}